# A New Scheme for Minimizing Malicious Behavior of Mobile Nodes in Mobile Ad Hoc Networks

Syed S. Rizvi and Khaled M. Elleithy

Computer Science and Engineering Department, University of Bridgeport, Bridgeport, CT, USA

{srizvi, elleithy}@bridgeport.edu

*Abstract-* **The performance of Mobile Ad hoc networks (MANET) depends on the cooperation of all active nodes. However, supporting a MANET is a cost-intensive activity for a mobile node. From a single mobile node perspective, the detection of routes as well as forwarding packets consume local CPU time, memory, network-bandwidth, and last but not least energy. We believe that this is one of the main factors that strongly motivate a mobile node to deny packet forwarding for others, while at the same time use their services to deliver its own data. This behavior of an independent mobile node is commonly known as misbehaving or selfishness. A vast amount of research has already been done for minimizing malicious behavior of mobile nodes. However, most of them focused on the methods/techniques/algorithms to remove such nodes from the MANET. We believe that the frequent elimination of such miss-behaving nodes never allowed a free and faster growth of MANET. This paper provides a critical analysis of the recent research wok and its impact on the overall performance of a MANET. In this paper, we clarify some of the misconceptions in the understating of selfishness and miss-behavior of nodes. Moreover, we propose a mathematical model that based on the time division technique to minimize the malicious behavior of mobile nodes by avoiding unnecessary elimination of bad nodes. Our proposed approach not only improves the resource sharing but also creates a consistent trust and cooperation (CTC) environment among the mobile nodes. The simulation results demonstrate the success of the proposed approach that significantly minimizes the malicious nodes and consequently maximizes the overall throughput of MANET than other well known schemes.**

## I. INTRODUCTION

Misbehavior in mobile ad-hoc networks occurs for several reasons. Selfish nodes misbehave to save power or to improve their access to service relative to others [1]. Malicious intentions result in misbehavior as exemplified by denial of service attacks. Faulty nodes simply misbehave accidentally. Regardless of the motivation for misbehavior its impact on the mobile ad-hoc network proves to be detrimental, decreasing the performance and the fairness of the network, and in the extreme case, resulting in a non-functional network [2]. This paper addresses the question of how to make network functional for normal nodes when other nodes do not route and forward packets correctly. Specifically, in mobile ad-hoc networks, nodes do not rely on any routing infrastructure but relay on packets for each other. Thus communication in mobile ad-hoc networks functions properly only if the participating nodes cooperate in routing and forwarding. However, it may be advantageous for in nodes not to cooperate, such as a selfish node wants to preserve own resource to save power, memory, network-bandwidth, and local CPU time. Therefore nodes assume themselves that other nodes would forward the packet. This selfish or malicious intention of nodes can significantly degrade the performance of mobile ad-hoc-networks by denial of service.

Many contributions to prevent misbehavior have been submitted so far, such as payment schemes for network services, secure routing protocols, intrusion detection, economic incentives and distributed reputation systems to detect and isolate misbehaved nodes. These exiting approaches alleviate some of the problems, but not all. In this paper, we focus on the design of a new time division based scheme that can avoid unnecessary elimination of malicious nodes while at the same time maximize the throughput of the system by increasing the recourse sharing among the mobile nodes. The existing methods/algorithms not only creating a performance bottleneck (i.e., by increasing the network congestion, transmission overhead etc.) but also diminishing the self-growing characteristic of a peer to peer network. These methods such as CONFIDANT [3] and CORE [4] force the participating nodes to adopt the same behavior as the other selfish nodes that have already been removed from the network due to the lack of resources. We believe that we should not propose any algorithm/method that becomes the reason for reducing the network resources and consequently force the existing participating nodes to behave exactly in the same way as other removed nodes. Instead, we strongly believe that we should come up with something that not only improves the resources and resource sharing but also creates a consistent trust and cooperation (CTC) environment among the mobile nodes.

The rest of this paper is organized as follows: Section II describes the research that has already been done in this area. The proposed analytical and mathematical models for CTC are presented in Section III. The simulation results are provided in section IV. Finally, section V concludes the paper.

## II. RELATED WORK

The terms *reputation* and *trust* are being used for various concepts in the literature, also synonymously [5, 6]. We define the term *reputation* here to mean the performance of a principal in participating in the base protocol as seen by others. By the term *trust* we denote the performance of a principal in the policing protocol that aims at protecting the base protocol.

The key thing in reputation system is watchdog and pathrater which have been proposed by Marti, Giuli, Lai and Baker [7]. They observed increased throughput in mobile ad-hoc networks by complementing DSR with a *watchdog* for detection of denied packet forwarding and a *path rater* for trust management and routing policy rating every path used, which enable nodes to avoid malicious nodes in their routes as a reaction. Their approach does not punish malicious nodes that do not cooperate, but rather relieves them of the burden of forwarding for others, whereas their messages are forwarded without complaint. This way, the malicious nodes are rewarded and reinforced in their behavior. They used a watchdog that identifies misbehaving nodes and a pathrater that helps routing protocols avoid these nodes. When used together in a network with moderate mobility, the two techniques increase throughput by 17% in the presence of 40% misbehaving nodes, while increasing the percentage of overhead transmissions from the standard routing protocol's 9% to 17%. During extreme mobility, watchdog and pathrater can increase network throughput by 27%, while increasing the overhead transmissions from the standard routing protocol's 12% to 24%.

CORE, a collaborative reputation mechanism proposed by Michiardi and Molva [4], also has a *watchdog* component; however it is complemented by a reputation mechanism that differentiates between subjective reputation (observations), indirect reputation (positive reports by others), and functional reputation (task-specific behavior), which are weighted for a combined reputation value that is used to make decisions about cooperation or gradual isolation of a node. Reputation values are obtained by regarding nodes as requesters and providers, and comparing the expected result to the actually obtained result of a request. Nodes only exchange positive reputation information.

A reputation-based trust management has been introduced by Aberer and Despotovic in the context of peer-to-peer systems [8], using the data provided by a decentralized storage method (P-Grid) as a basis for a data-mining analysis to assess the probability that an agent will cheat in the future given the information of past transactions.

A context-aware inference mechanism has been proposed by Paul and Westhoff [9], where accusations are related to the context of a unique route discovery process and a stipulated time period. A combination is used that consists of un-keyed hash verification of routing messages and the detection of misbehavior by comparing a cached routing packet to overheard packets.

The EigenTrust mechanism was proposed by Kamvar, Schlosser and Garcia-Molina [10] which aggregates trust information from peer by having them perform a distributed trust calculation approaching the Eigenvalue of the trust matrix over the peers. The algorithm relies on the presence of pre-trusted peers, that is some peers have to be trusted, prior to having interacted with them. By isolating peers with bad reputation, the number of inauthentic downloads is decreased, however, if the motivation for misbehavior is selfishness, the misbehaved peers are rewarded. If the download is not successful, the peer is removed from the list of potential downloads. A potential drawback of this approach is that it provides an incentive to change one's identity after having misbehaved. A formal model for trust in dynamic networks based on a policy language has been proposed by Carbone, Nielsen, and Sassone [11]. They express both trust and the uncertainty of it as *trust ordering* and *information ordering*, respectively. In their model, only positive information influences trust, such that the information ordering and the trust ordering can differ. In our system, however, both positive and negative information influence the trust and the certainty, since we prefer $p$ positive observations that come out of $n$ total observations to $p$ out of $N$ when $n<N$.

## III. PROPOSED ANALYTICAL AND MATHEMATICAL MODELS FOR CREATING CONSISTENT TRUST AND COOPERATION

This section first presents an analytical model that gives our hypothesis to mitigate the problem of misbehavior among the mobile nodes. Secondly, we use the proposed analytical model to create a corresponding mathematical model. The creation of mathematical model can be viewed as a formalization of the proposed hypothesis. Based on the proposed mathematical model, we perform the numerical and simulation analysis for variety of scenarios in two parts. First, we use the mathematical model to run different scenarios in order to determine the performance of Ad-hoc networks by analyzing different critical network parameters such as throughput, transmission overhead and the utilization. Secondly, we use the same set of parameters as a performance measure.

### A. The Proposed Analytical Model

We model the Ad-hoc network in much the same way as other researcher does except this paper introduces the new concept of time division. The idea of time division can simply be envisioned by considering a particular node of a network that has a potential to misbehave in the absence of the sufficient resources require to forward the packets of the neighboring nodes. This implies that if one can ensure that the network has enough resources that can be shared equally among the network nodes, then it can be assumed that the possibility of node misbehavior degrades significantly. Thus this reduction in the node misbehavior can be achieved through the time division technique that divides the time asymmetrically into the following two times: transmission-time required for *node-packets* and transmission-time required for *neighbor-packets*. The asymmetric division enables a node to effectively adjust the time required to transmit its own packets and/or the neighbor's packets. The reason for using the asymmetric division of the available time is to allow a node to effectively utilize the time by dividing it with respect to its current status (i.e., the available recourses) and consequently utilizing the bandwidth in an efficient manner. The efficient utilization of the bandwidth satisfies the requirement of the fairness which is one of the key factors that forces a node to unfair with its neighbor. This indirectly points that we reduce the chances of misbehave since the node now has a total authority on the available resources. It should also be noted that we adopt an asymmetric approach to work with the time division method for this research which opposed to the conventional division of time (i.e., the symmetric or equal division employed by many different techniques).

Thus this clearly allows a node to optimize the use of network parameters such as throughput, transmission overhead and the utilization by effectively utilizing the total time with respect to the current situation of the network. In other words, the proposed hypothesis can be considered as a dynamic mechanism that allows all nodes to perform performance optimization at run time by intelligently using the available time which is one of key elements of any system. In either case, the proposed hypothesis moves the control from the resources to nodes.

## B. The Proposed Mathematical Model

Before going to develop the actual mathematical model based on the above analytical model, it is worth mentioning some of our key assumptions. These assumptions help understanding the complex relationship between a large numbers of parameters. For the proposed mathematical model, we assume that a system has $K$ nodes where each individual node $k$ not only works as a normal mobile station but also works as a packet forwarding device for the other nodes. In addition, we assume that any kind of topology can be implemented among the mobile nodes to construct the Ad-hoc network. This assumption allows us to implement different scenario (such as a node can have any number of input and output lines) on each node of the network to show the consistency of the proposed analytical and mathematical model. For the ease of simplicity, we perform the numerical analysis for a single node $k$. This can be further extended for the whole network by computing the collective behavior of the Ad-hoc network. This approach allows a reader to grape the idea of the proposed method from a very simple equation to highly complex derivations. The systems parameters along with their definitions are listed in Table I.

TABLE I. SYSTEM PARAMETERS DEFINITIONS

| Parameters | Definition |
|---|---|
| $t_i$ | Total time use by node |
| $t_p$ | Time that spend node on personal packets |
| $t_{np}$ | Time that spend node on neighbor packets |
| $T_{put}$ | Through put of the node |
| $D_R$ | Data rate on route |
| $K$ | Is the no of packets |
| $N_p$ | Node power |
| $N_{pp}$ | Node power use on personal packets |
| $N_{np}$ | Node power use on neighbor packets |
| $K_{pout}$ | Personal Packets that goes out form node |
| $K_{nout}$ | Neighbor packets that goes to out from node |
| $K_{nin}$ | Neighbor packets that come in the node |
| $U_t$ | Total utilization |
| $U_{R_n}$ | Utilization on number of route |

The primary principal of Ad-hoc network is that it allows each node of the network to fully participate in the construction of the network. The word *fully participation* leads us to the fact that a node not only transmits its own packets to the other neighboring nodes but also provides its services to other nodes as a forwarding device. For the proposed method, we assume that a node can decide to transmit its own packets with a certain probability while at the same time it can also deny the transmission of the other neighboring packets with a difference of a certain probabilities. In simple words, we can develop a relationship between these two probabilities as follows: a node can transmit the self generated packet(s) with a probability of p where as it can transmit its neighbor packet(s) with the probability of q.

Suppose, *p* is the probability for which a node forwards personal packets where as *p (1 – p)* is the probability for which a node transmit packets received from one or more neighbors. In addition, we assume that *k* is total number of packets that can be transmitted by a certain node of the Ad-hoc network. The total numbers of packets include both the self generated packets and the packets receive from one or more nodes. Taking this into account, we can say that if the probability of transmission of a single packet is *(1-p)ˣ* where *x* represents a single packet, then the probability to transmission *k* packets would be *(1-p)ᵏ* where *k* represents the total number of packets that a node can transmit. This leads us to the following mathematical fact:

$$\left(1-p\right)^{k} \qquad (1)$$

Equation (1) can simply be formalized for *k* number of packets as follows:

$$p\left(1-p\right)^{k} \qquad (2)$$

As mentioned earlier, the proposed method is exclusively dependent on the time division methodology where a node can divide the time asymmetrically to represent the time it needs to transmit self generated packets as well as the time it takes to transit the packets arriving from one or more nodes. To make our proposed approach more realistic, we assume that if the packet that resides in a certain node is not delivered to its intended destination within the specified time, then that packet must be discarded by the node. The lost of the packet at the node level forces us to retransmit the packet. This assumption is essential for us to make our derivations close to what actually happen in the real world. This also helps us demonstrating the effectiveness of the proposed algorithm in the presence of packet retransmission.

For the ease of understating, we assume that the time a node takes to transmit self generated packet can be represented as $t_{pp}$ where as the time it takes to forward the packets received from one or more neighbors is represented as $t_{np}$. It should be noted that the total available time per node is just the sum of the time a node takes to transmit self generated packet and time it takes to forward the packets received from one or more neighbors. This relationship can be mathematically expressed in the following equation:

$$t_{i} = t_{pp} + t_{np} \qquad (3)$$

where *i* represents the index of node that can be expended from 1 to *K* (i.e., *K* represents the total nodes present in a Ad-hoc network)

As we mentioned in the introduction that there are some critical parameters such as throughput, transmission overhead and the utilization that one should consider when the intention is to perform a true evaluation of a network. Based on this, we are now in the position to give our hypothesis about one of the key parameter, system throughput. The maximum throughput is defined as the asymptotic throughput when the load (the amount of incoming data) is very large. In packet switched network where the load and the throughput are equal, the maximum throughput may be defined as the load in bits per seconds. Thus this in turns lead us to a fact that the maximum throughput can not be defined in the presence of packet drops at the node level. As mentioned earlier, to make our model more realistic we consider the possibility of packet drops and consequently the packet retransmission at the node level. In addition to that, we believe that the maximum throughput can only be defined when the delivery time (that is the latency) asymptotically reaches infinity. The second argument is absolutely not true for the proposed algorithm, since we have a finite time available per node that indicates the presence of finite bandwidth. That is both of them are the realistic assumptions made by us for proof the authenticity of the proposed time division technique. Thus these two arguments force us to derive a new formula that behaves with respect to the proposed time division technique. The throughput from the proposed algorithm for a certain node of the Ad hoc network can be computed as follows:

$$T_{put} = \frac{Total\ Packets\ Forwarded}{Total\ Time} \qquad (4)$$

The denominator of (4) is derived from (3) where as the numerator of equation is determined by using (1) and (2). One can see that as we increase the left hand side of (2), it causes a decrease in the left hand side of (4). It should also be noted that as we increase the sum of (1) and (2), it significantly increases the left hand side of (4). To make these relationships simple, we can say that the increase in the sum of (1) and (2) causes an increase in the throughput where as an increase in the total time that is determined by (3) causes a decrease in the throughput per node. This is because the more we increase the time, the more bandwidth we need to reserve to satisfy the transmission requirements.

A significant increase in the bandwidth utilization (which is beyond the scope of the available bandwidth per node) represents degradation in the throughput that indicates an increase in the possibility of node misbehavior. Thus, this implies that the proposed algorithm is not only improving the performance but also providing a chance to choose the optimal values of critical parameters (such as time) to achieve comparatively better performance than the others well known Ad- hoc networks routing algorithms. Equation (4) can be further simplified in the following form:

$$T_{put} = \frac{Node's\ Packets + Neighbour's\ Packets}{Total\ Time} \qquad (5)$$

To formalize the above discussion, we can combine probabilities of transmission from (1) and (2) with the total available time per node from (3) in (5). Thus this expresses the node throughput not only by means of total available time but also by means of the total number of packets a node can transmit. The final result can be expressed in the following equation:

$$T_{put} = (1-p)^k + (1-p)^k / t_i \qquad (6)$$

It should be noted that (6) gives node throughput by considering the time $t_i$ spends on a single packet (that is the time spend on one packet is the sum of the time spend on self generated packets and the neighbor packets). Solving (5) for $k$ number of packets in terms of the total time required by a node can be expressed in the following equation:

$$t_i = \sum_1^k t_{pp(k)} + \sum_1^k t_{np(k)} \qquad 1 \le k < \infty \qquad (7)$$

where $k$ in (7) represents the number of packets that are bounded between 1 and the infinity. The first and the second quantity of the right hand side of (7) are indicating the time required transmitting the self generated packets and the time required to transmit the neighbor packets. In addition to that, it would be interesting to compute the time that the node can spend in transmitting the self generated packets and compare it with the time required to transmit the neighbor packets. For doing this, one may need to generate a generic time that can be further used in computing the specific time. The generic time equation can simply be stated as:

$$t = \frac{no\ of\ packet}{data\ rate} \qquad (8)$$

Using (8), one can now compute the two major components of the proposed time division algorithm. It is essential in order to understand the concept of asymmetric division. One of the two asymmetric time division quantities can be quantified as follows:

$$t_{np} = \frac{P(1-P)^k}{D_R} \qquad (9)$$

where $D_R$ in (9) represents the data rate.

Recall one of our fundamental assumptions that a node transmits $k$ number of packets in total time $t_i$. This assumption allows us to set up a lower and upper bound on the number of packets that a node can transmit. Therefore, the limit for $k$ should exist somewhere zero to infinity. One of the main reasons for recalling this assumption is make a more generalized form of (9). That is we need to derive the same expression for $k$ number of packets that a node can transmit. In addition to this assumption, let $t_{np}$ is the time taken by a node to forward packets received from one or more neighbors. Taking these two factors into account, one can generalize (9) as follows:

$$t_{np} = \sum_{k \ge 1}^{k \le \infty} \frac{P(1-P)^k}{D_R} \qquad where\ 1 \le K \le \infty \qquad (10)$$

The numerator of (10) is just a summation of total packets forwarded by a node with respect to the probabilities set up at static time. Similarly, the denominator is the data rate at which the numbers of bits per packets are arrived at the destination (note that the destination in this case is the targeted node). One of the main advantages of this generalization is the analysis of the proper behavior of a node in the presence of malicious node.

In similar manner, one can derive the corresponding generalized form of an equation for node's personal packets. This introduction, therefore, allows us to make the following assumption and derive another mathematical expression for node's personal packets. If $t_{pp}$ is the total time taken by a node to forward its own $k$ number of packets, then equation for $t_{pp}$ can be rewritten as.

$$t_{pp} = \sum_1^k \left\{ \frac{(1-P)^k}{D_R} \right\} \qquad where\ 1 \le K \le \infty \qquad (11)$$

Equation (11) is the summation of probabilities of one packet to k number of packets per node in the presence of a certain data rate. The numerator of (11) is just a summation of total packets forwarded by a node with respect to the probabilities set up at static time. Similarly, the denominator is the data rate at which the numbers of bits per packets are arrived at the destination (note that the destination in this case is the targeted node). It should be noted that the same proposed equations will be used to conduct the analysis of the proposed mathematical model.

In order to extend our proposed mathematical model, one needs to derive an expression for the throughput per node. To follow the same bottom-up mathematical technique, we need to proceed from one node to $n$ number of nodes. We begin the derivation for throughput by recalling one of our fundamental equations of total time taken by a single node. By substituting the value of total time $t_i$ from (3) into (6), we get

$$T_{put} = \frac{\left\{ (1-p)^k + p(1-p)^k \right\}}{\left\{ t_{pp} + t_{np} \right\}} \qquad (12)$$

In order to generalize (12), we need to substitute the values of $t_{pp}$ and $t_{np}$ from (10) and (11), respectively, into (12), we get:

$$T_{put} = \sum_{1}^{k} \left[ \frac{D_R\left\{(1-p)^k + p(1-p)^k\right\}}{(1-p)^k + p(1-p)^k} \right] \qquad (13)$$

The first two quantities in denominator of (13) represent the summation of the time a node takes to transmit the personal packet and the neighbor's packets. Where as the numerator is the summation of probabilities set up for both the personal packets and the neighbor packet. It should be noted that (13) is generalized in a sense that it accommodates k number of packets that a node can deal at a certain point of time. To make it simple, we can rewrite equation as follows:

$$T(put\ of\ node) = \sum_{1}^{k} \left[ \frac{D_R \times (1-p)^k + p(1-p)^k \times D_R}{\left\{(1-p)^k + p(1-p)^k\right\}} \right] \qquad (14)$$

Equation (14) is the total throughput of a node for $k$ number of packets that a node can transmit (i.e., both the personals packets and the neighbor's packets). For a small set of numerical analysis using the mathematical expressions derived above, let the value of k =1. This leads us to the following result: $T_{put} = D_R$. This result can be interpreted by understanding different conditions and/or assumptions. For instance, if we assume that a node becomes selfish, consequently it does not forward the packets which were received from one of its neighboring nodes.

Based on the above analysis, one can conclude that the throughput of the system depends mainly on the factors or parameters that we include in different equations of the proposed mathematical model. Increasing or decreasing the values of these parameters result in different performance from node to node. However, it would be more interesting to account those parameters that are not directly related to the internal components of a node. One of the best ways to consider these parameters is to compute the utilization per node and extend the derived mathematical expression for typically $n$ number of nodes. It is expected that the utilization of node remains stable as long as the node utilizes the available route efficiently. However, the utilization per node may degrade due to the improper use of available channels. Thus, this clearly shows that we need to consider the node utilization per channel and need to extend that expression for generalization. To make this practical, let us assume that $Np$ is the power of node and K is the number of packet that a node can transmit. Taking these assumptions into account, one can derive a generic expression for utilization as follows:

$$U = \frac{N_{pout}}{N_{pin}} \qquad (15)$$

We call (15) as a generic mathematical expression of utilization, since both the numerator and the denominator are unknown and need to be determined to find out a more specific expression. Before going to utilize a bottom-up methodology, it is worth mentioning that node power is distributing non-uniformly among the packets almost in the same way as we distribute the time. Therefore, this new concept of power division leads us to the following mathematical expression for node-utilization with respect to the node's personal packets.

$$N_{ppout} = \sum_{1}^{K} \left\{ \frac{K_{Pout}}{t_{pp}} \right\} \qquad (16)$$

It should be noted that (16) is a more specific form of (15) since it only account for the personal packets. In addition to that, it can be considered a generalized form since it includes a large number of packets whose value may vary from one to infinity. To make this model equivalent, one can derive the same expression to compute the utilization per node that is related to the packets receive by the targeted node from one of its neighboring nodes. Thus the opposite hypothesis leads us to the following mathematical expression for the node utilization with respect to the personal packets:

$$N_{pnout} = \sum_{K \geq 1}^{K < \infty} \left\{ \frac{K_{nout}}{t_{np}} \right\} \qquad (17)$$

Contrary to (17), there should be an equivalent possibility of node inputs that can easily be computed as follows:

$$N_{pnin} = \sum_{1}^{K} \left( \frac{K_{nin}}{t_{np}} \right) \qquad (18)$$

It should be noted that (18) can be useful to compute the output of the nodes in terms of the inputs of the node. In other words $N_{Pout}$ is the sum of work on outgoing personal and neighbor packets that lead us to derive the simple mathematical relationship:

$$N_{pout} = N_{pp(out)} + N_{pnout} \qquad (19)$$

In order to show that (19) is a valid true mathematical relationship between the input and output lines of a node, one needs to give another relationship as follows:

$$N_{pin} = N_{pnin} \qquad (20)$$

This should now be clear that one of the reasons for deriving the above two relationship is to derive a more general expression from (16) and (17). Therefore, by substituting (16) and (17) into (19), we get the following equation:

$$N_{ppout} = \sum_{k \geq 1}^{k < \infty} \left[ \left\{ \frac{K_{ppout}}{t_{pp}} \right\} + \left\{ \frac{K_{nout}}{t_{np}} \right\} \right] \qquad (21)$$

Similarly, we can derive another expression using (20) which opposed to (21) as follows:

$$N_{Pin} = \sum_{1}^{K} \left( \frac{K_{nin}}{t_{np}} \right) \qquad (22)$$

The last two equations (i.e., (21) and (22)) can now be used to derive the final expression for utilization as follows:

$$U = \sum_{1}^{k} \frac{\left\{ \frac{K_{pout}}{t_{pp}} \right\} + \left\{ \frac{K_{nout}}{t_{np}} \right\}}{\left\{ \frac{K_{nin}}{t_{np}} \right\}} \qquad (23)$$

All lines that are used for transferring the data or packets are also used for receiving the data or packets from neighbor nodes. This implies that the utilization per channel or line can be computed using (23). If we denote this line-utilization as (24), we can extend it to generalized (23).

$$U_R = \sum_{k \geq 1}^{k < \infty} \frac{\left( K_{pout} / K_{nout} \right) t_{np}}{K_{nin}} \qquad (24)$$

If we assume that $n$ numbers of routes are attached through the targeted node, then the utilization of the targeted node on all routes can simply be computed by summing the utilization of each node per channel. In other words, (24) needs to be run and sum for $n$ numbers of routes that are connected to a certain node. This can lead us to the following equation:

$$U_t = \sum_{n \geq 1}^{n < \infty} U_{R_n} \quad \text{where } 1 \leq n < \infty \qquad (25)$$

This can also be interpreted as follows:

$$U_t = U_{R1} + U_{R2} + U_{R3} + \dots\dots + U_{Rn} \qquad (26)$$

Therefore, the total utilization of system can be derived from (23) and (25) as follows:

$$U_t = \sum_{n \geq 1}^{n < \infty} \sum_{k \geq 1}^{k < \infty} \frac{\left( K_{pout} / t_{pp} \right) + \left( K_{nout} / t_{np} \right)}{K_{nin} / t_{np}} \qquad (27)$$

We perform some simplification in (27) that results the following equation:

$$U_t = \sum_{n \geq 1}^{n < \infty} \sum_{k \geq 1}^{k < \infty} \frac{1}{K_{nin}} \left[ K_{pout} \left( t_{np} / t_{pp} \right) + K_{nout} \right] \qquad (28)$$

The above equation can be used to compute the total utilization of a certain node for all packets that it can forward and/or receive from one of its neighbor though all possible channels.

## IV. EXPERIMENTAL VERIFICATIONS AND PERFORMANCE ANALYSIS OF CTC SCHEME

We have shown that the system throughput can be measured in term of packets that neighboring node is generated as well as the self generated packets. When it comes to performance, it is a standard in a wireless Ad-Hoc network to determine the performance of the network in terms of node misbehavior by looking at the ratio between the packet drop per node and the total number of packets per node. In other words, in order to quantify the node-misbehavior that apparently looks a philosophical concept one can need to compute that how much packets the node is dropping that it should forward to the intended destination. To make the proposed methodology up to the standard, we derive the formula for computing the packet drop per node using (5).

As mentioned earlier, we determine the behavior of the malicious node in terms of the number of packets that should have transmitted to the intended destination. For taking this into account, one can say that the effective throughput of a node is entirely dependence on how efficiently the node is forwarding the neighbor packets and thus creating a consistent trust environment among the nodes. This argument, therefore, allows us to make minor changes in (5).

$$PacketDrop = \frac{\left( NodesPacket \right) + \left( NeighborPacket \right)}{TotalTime}$$

For the ease of clarity, we make some implicit assumptions that remain same for all the investigated algorithms presented in our simulation results. These include an initial small probability of fixed packet drop that remain same for all algorithms. The reason for making an initial value of packet drop as an assumption is due to the fact that we are unaware that how the nodes misbehave when they first boot up. Instead of considering this value as zero, it would rather useful to smooth out the effects that result due to the malicious behavior of nodes. Thus, this significantly clarifies the performance difference between the proposed and the other well known techniques.

### A. Case I

Before discussing the simulation results, it is worth mentioning some of our key assumptions that we made for the sake of experimental verification for the proposed CTC algorithm. Some of them are as follows. For case-1 we assume that the self generated packets per node is constant (i.e., node generates a fixed number of packets for a specified amount of time that remains same for both CTC and DSR algorithms). We assume that one of the neighboring nodes of the target node sends packets at a certain rate that will increase linearly over the total simulation time. This assumption helps understanding the true performance of the proposed CTC algorithm.

Fig. 1 shows the simulation results of packet-drops per node with respect to the number of packets generated by one of the neighboring nodes. It should be noted that as we increase the self generated packets, the number of packet-drops per node is increased. In addition, it can be seen in Fig. 1 that for a small value of neighbor packet generation typically 500, both CTC and DSR are overlapping each other. However a slight increase in the neighbor packet generation causes a performance difference between these two approaches. In other words, an increase in neighbor packet generation forces the DSR to perform poorly as compared to the proposed CTC algorithm. Thus the node-misbehave increases for the DSR algorithm whereas it gives a consistent behavior for the proposed CTC algorithm. Fig. 1 suggests that for large value of neighbor packet generation (typically after 800 to 1600), the proposed CTC algorithm successfully maintain a consistent node misbehavior (typically the node misbehavior for the proposed CTC algorithm exists between 20 to 25 percent) where as the node misbehavior increases linearly in the case of DSR algorithm. Based on the results (Fig. 1), one can say that the proposed CTC algorithm outperforms the DSR algorithm for a large neighbor packet generation.

### B. Case II

The CASE-II of our simulation is different from CASE-I in such a way that now both inputs of a node-forwarding system become a linear function of the node-time. In other words, for CASE-II we are not only increasing the neighbor-generated packets but also increasing the self-generated packets. The simulation result of this case satisfies the proposed mathematical model discussed in Section III in a way that the overall packet drop performance of both investigated algorithms decreases. In other words, it can be seen that the packet drop is more rapid in Fig. 2 with respect to the neighbor-generated packets.

In harmony with our expectations, as the number of neighbor-generated packets increased, the packet-drop performance of the proposed algorithm degraded. However, the performance degradation of the proposed algorithm was small compared to the performance degradation of the DSR algorithms. The packet-drop performance of the CTC algorithm below 40 neighbor-generated packets is almost similar to that of the CTC algorithm as shown in figure of CASE-II. However, the amount of the packet-drop improvement for the proposed CTC algorithm over the DSR algorithm increases with respect to the values of neighbor-generated packet.

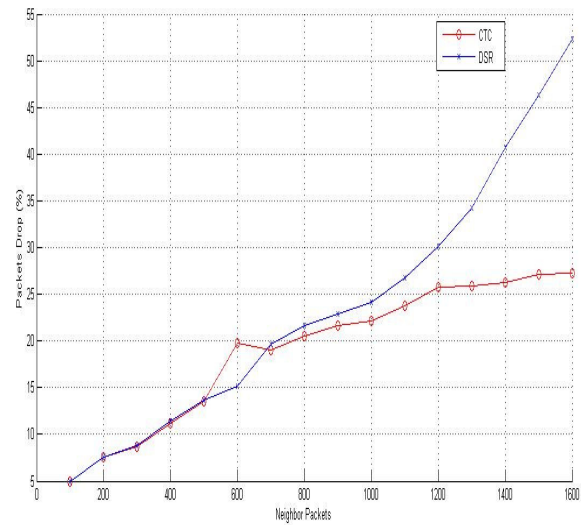

Figure 1. Neighbor packet generation vs. packet drop

### C. Case III

The parameters-assumption for CASE-III is different from the previous cases in such a way that now one input (that is the neighbor-generated packets) of a node-forwarding system becomes a linear increasing function of the node total time where as the input (that is the neighbor-generated packets) becomes a linear decreasing function of the node total time. In other words, for CASE-III we are interested to see the packet drop performance of the investigated algorithms (that is the proposed CTC algorithms as well as the DSR algorithm) in the presence of both increasing and decreasing functions. The expected output of this simulation was exactly the same as we were expecting based on our proposed mathematical model. That is the values of packet-drop for both CTC and DSR decreases as compared to the other two cases we discussed above. This is due to the fact that we consider the number of self-generated packet as a decreasing linear function of the node total time while at the same time we use the neighbor-generated packets as an increasing function as shown in Fig. 3.

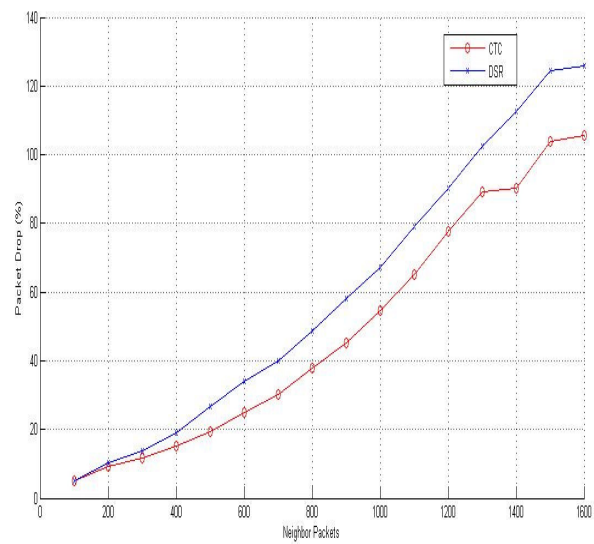

Figure 2. Neighbor packet generation vs. packet drop

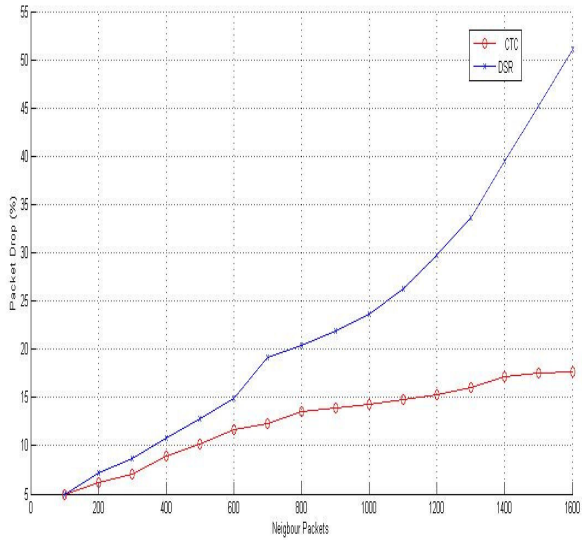

Figure 3. Neighbor packet generation vs. packet drop

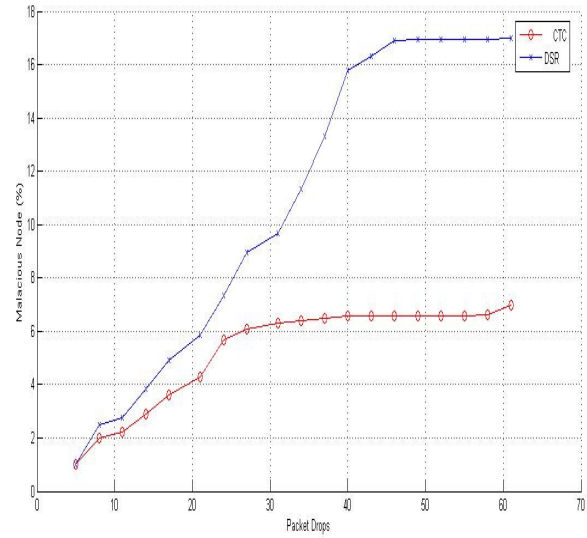

Figure 5. Malicious Nodes (%) vs. packet drop

### D. Case IV

CASE-IV is yet another verification of the proposed mathematical model. For this case, we assume that the neighbor-generated packets is a constant function of time (i.e., we use a constant value for this system parameter and used it with respect to time throughput the simulation of

CASE-IV). On the other hand, we consider self-generated packets as a linear increasing function of the total node time. It should be noted that the term linear increase or decrease implies a constant uniform change in the system parameter with respect to time. This case can also be considered as a reciprocal of CASE-I from its fundamental assumptions point of view. Thus we should also expect a reciprocal output for this simulation (that is its packet drop performance should behave exactly the opposite as we have seen in Fig. 1). With harmony to our expectations, the packet drop remains constant for all values of neighbor packets as shown in Fig. 4.

### E. Case V

This case describes the effect of the last four cases in terms of node malicious behavior as shown in Fig. 5 and 6. We used the statistics to derive a relationship between the node malicious behavior and the ratio packet drop. As one can see that in both Fig. 5 and 6, the number of malicious node becomes an increasing function of the packet drop performance which also validates the structure of our proposed mathematical model. However, the performance differences between the two investigated algorithms from malicious nodes perspective are quite subtle. That is a less number of nodes misbehave in the case of the proposed CTC algorithm when compared to the DSR algorithms. For small value of packet drops typically 10, both algorithms are overlapping each other but however, as the number of packet drop increases, the proposed algorithms giving much better performance than the other algorithms.

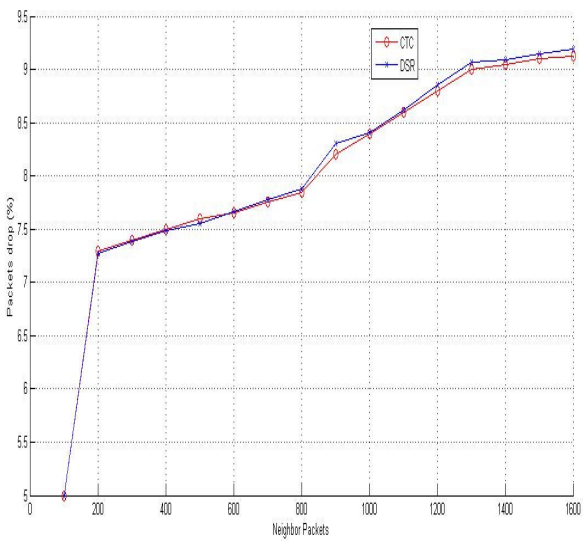

Figure 4. Neighbor packet generation vs. packet drop

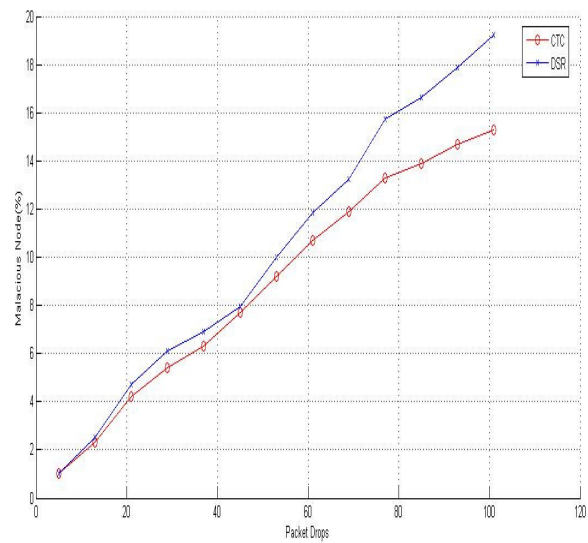

Figure 6. Malicious Nodes (%) vs. packet drop

## V. CONCLUSION

This paper presented a critical analysis of the recent research wok and its impact on the overall performance of a mobile Ad hoc network. We provided a discussion on some of the common misconceptions in the understating of selfishness and miss-behavior of nodes. Moreover, this paper proposed both analytical and mathematical model that can be used to effectively reduce the number of malicious nodes and packet drops. Our simulation results demonstrated that the proposed mathematical model not only points out the weaknesses of the recent research work but also approximates the optimal values of the critical parameters (such as the throughput, transmission over head, channel capacity and utilization etc.) that have great impact on the overall performance of a mobile Ad hoc network. Simulation results presented in this paper show that how the performance of mobile Ad hoc networks degrades significantly when the nodes eliminations are frequent. The simulation results of this paper are completely based on the proposed mathematical model for both lightly and heavily loaded networks. These results addressed many critical system parameters such as packet drop and packet loss versus malicious nodes, neighbor packet generation and drop ratio, and throughput per node per system. Our simulation study is also a comparison with the most recent and well admitted research work such as CONFIDANT and CORE. This comparative study provides a proof of our proposed methodology that appears to be correct.

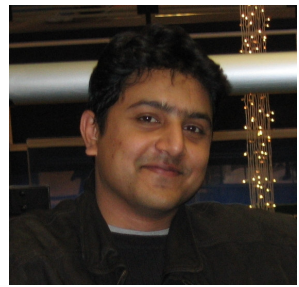

**Syed S. Rizvi** is a Ph.D. student of Computer Science and Engineering at University of Bridgeport. He received a B.S. in Computer Engineering from Sir Syed University of Engineering and Technology and an M.S. in Computer Engineering from Old Dominion University in 2001 and 2005, respectively. In the past, he has done research on bioinformatics projects where he investigated the use of Linux based cluster search engines for finding the desired proteins in input and outputs sequences from multiple databases. For last three year, his research focused primarily on the modeling and simulation of wide range parallel/distributed systems and the web based training applications. Syed Rizvi is the author of 68 scholarly publications in various areas. His current research focuses on the design, implementation and comparisons of algorithms in the areas of multiuser communications, multipath signals detection, multi-access interference estimation, computational complexity and combinatorial optimization of multiuser receivers, peer-to-peer networking, network security, and reconfigurable coprocessor and FPGA based architectures.

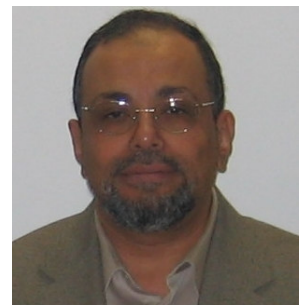

**Khaled Elleithy** received the B.Sc. degree in computer science and automatic control from Alexandria University in 1983, the MS Degree in computer networks from the same university in 1986, and the MS and Ph.D. degrees in computer science from The Center for Advanced Computer Studies at the University of Louisiana at Lafayette in 1988 and 1990, respectively. From 1983 to 1986, he was with the Computer Science Department, Alexandria University, Egypt, as a lecturer. From September 1990 to May 1995 he worked as an assistant professor at the Department of Computer Engineering, King Fahd University of Petroleum and Minerals, Dhahran, Saudi Arabia. From May 1995 to December 2000, he has worked as an Associate Professor in the same department. In January 2000, Dr. Elleithy has joined the Department of Computer Science and Engineering in University of Bridgeport as an associate professor. Dr. Elleithy published more than seventy research papers in international journals and conferences. He has research interests are in the areas of computer networks, network security, mobile communications, and formal approaches for design and verification.